\documentclass[runningheads]{llncs}

\usepackage{dialogue}
\usepackage{tabularx}
\usepackage{float}
\usepackage{amsmath}
\usepackage{subcaption} 
\usepackage[T1]{fontenc}
\usepackage{graphicx}
\usepackage{multirow}
\usepackage{enumitem}
\usepackage{xcolor}
\usepackage{ltablex}

\usepackage{longtable}
\usepackage{multirow}
\keepXColumns

\begin{document}
\title{Does the \textit{TalkMoves} Codebook Generalize to One-on-One Tutoring and Multimodal Interaction?}

\titlerunning{Does \emph{TalkMoves} Generalize to One-on-One Tutoring?}

\author{
Corina Luca Focsan\orcidID{0009-0005-7427-9301}\inst{1} \and
Marie Cynthia Abijuru Kamikazi\orcidID{0009-0006-6586-9497}\inst{1} \and
Tamisha Thompson\orcidID{0000-0002-0155-6540}\inst{2} \and
Jennifer St. John\orcidID{0000-0002-0248-8586}\inst{2} \and
Kirk P. Vanacore\orcidID{0000-0003-0673-5721}\inst{2} \and
Danielle R. Thomas\orcidID{0000-0001-8196-3252}\inst{1} \and
Kenneth R. Koedinger\orcidID{0000-0002-5850-4768}\inst{1} \and
René F. Kizilcec\orcidID{0000-0001-6283-5546}\inst{2}
}

\authorrunning{Luca Focsan et al.}

\institute{
Carnegie Mellon University\\
\email{\{clucafoc, mabijuru, dchine, kk1u\}@andrew.cmu.edu}
\and
Cornell University\\
\email{\{js3953, tst47, kpv27, kizilcec\}@cornell.edu}
}
\maketitle 

\begin{abstract}
Accountable Talk theory has been widely adopted to analyze classroom discourse and is increasingly used to annotate tutoring interactions. In particular, the \emph{TalkMoves} codebook, grounded in Accountable Talk theory, is commonly used to label tutoring data and train models of effective instructional support. However, Accountable Talk was originally developed to characterize collaborative, whole-classroom oral discourse, \emph{not} identify talk moves in one-on-one tutoring environments using multimodal data (e.g., video, audio, chat). As tutoring platforms expand in scale and modality, questions remain about whether Accountable Talk-based codebooks generalize reliably beyond their original classroom context and data representation. This study examines whether the human-developed \emph{TalkMoves} codebook generalizes in reliability, utility, and interpretability when applied to one-on-one tutoring across audio, chat, and multimodal data. We compare \emph{TalkMoves} with a hybrid AI-human developed codebook using a workflow established in prior research. Two expert annotators with over 20 years of teaching experience applied both codebooks to six tutoring sessions spanning three modalities: chat-based, audio-only, and multimodal interactions. Results show that while \emph{TalkMoves} achieved higher overall inter-rater reliability than the AI-human codebook ($\kappa$ = 0.74 vs. 0.64), the AI-human codebook demonstrated broader empirical coverage and higher perceived usability across modalities. Both codebooks undercaptured tutoring-relevant moves and introduced ambiguity when identifying actions expressed through non-verbal and multimodal artifacts. Together, these findings highlight the uneven generalizability of \emph{TalkMoves} to tutoring contexts and motivate the development of modality-aware, tutoring-grounded codebooks.

\end{abstract}

\section{Introduction}
Qualitative coding is a core method for analyzing learning processes in instructional discourse by defining conceptual categories and identifying their occurrence to uncover themes \cite{gibbs2018analyzing}. Despite its value, this process is time-consuming and labor-intensive \cite{castleberry2018thematic}. As a result, researchers analyzing  tutoring data, which is increasingly multimodal, often rely on established codebooks such as \textit{TalkMoves}, initially developed with classroom dialogue \cite{michaels2008deliberative,suresh2021using}. This introduces two key gaps: a contextual mismatch between tutoring and classroom settings, and a representational mismatch between multimodal data and oral discourse. Prior work shows that classroom-oriented codes are rarely used in tutoring, while tutoring-specific functions are undercaptured \cite{balyan2022modeling,booth2024human}. What remains underexplored is how well the \textit{TalkMoves} codebook generalizes across tutoring data modalities.

This work examines how Accountable Talk-based codebooks generalize across tutoring data modalities. We compare the \textit{TalkMoves}\footnote{https://doi.org/10.48550/arXiv.2204.09652} codebook with a hybrid AI-human codebook developed using a validated workflow \cite{barany2024chatgpt}, to understand how context and data modality shape codebook performance. Two expert annotators applied both codebooks to tutoring transcripts across chat, audio, and multimodal modalities. We investigate: (\textbf{RQ1}) To what extent do the human-developed \textit{TalkMoves} codebook and AI-human developed codebook maintain \textbf{reliability} and \textbf{utility} across one-on-one tutoring interaction data? (\textbf{RQ2}) What \textbf{themes exist when comparing and contrasting} \textit{TalkMoves} and the AI-human codebook, prompted using tutoring data?

\vspace{-2mm}
\section{Related Work}
\vspace{-1mm}
\textbf{Accountable Talk Theory and the \textit{TalkMoves} Codebook}. Accountable Talk theory defines discourse practices that support collective knowledge construction through structured dialogue, with strong evidence linking rich instructional talk to improved learning outcomes \cite{michaels2008deliberative,webb2019teacher}. Central to this framework are \textit{talk moves}, discourse actions used by teachers and students to elicit reasoning \cite{michaels2010accountable}. These moves are organized around accountability to the learning community, content knowledge, and rigorous thinking, and have been widely adopted across instructional practice \cite{o2019supporting}. The \textit{TalkMoves} codebook operationalizes this framework through a set of teacher and student moves designed for reliable annotation of classroom discourse, though it is not intended to be exhaustive \cite{suresh2021using}. 

\textbf{LLMs for Codebook Development.} The time- and labor-intensive nature of qualitative codebook development has led to growing interest in using large language models (LLMs) to support this process. Barany et al. \cite{barany2024chatgpt} compared multiple human and AI workflows, finding that hybrid approaches, where LLMs generate initial codes followed by human refinement, achieve higher inter-rater reliability and utility ratings. However, these workflows are largely inductive, raising concerns that LLM-generated codes may reflect non-neutral or scientifically misaligned constructs \cite{nguyen2025misrepresentation}. To address this, Zambrano et al. \cite{zambrano2026data} incorporate theory into LLM prompting, demonstrating that theory-informed codebooks are  more theoretically aligned and more usable for domain experts.
\vspace{-2mm}
\section{Methods}
\textbf{Data Source.} We analyzed mathematics tutoring data from two nonprofit organizations across chat-based, audio, and multimodal data modalities. Chat data came from an on-demand platform providing free, 24/7 tutoring to U.S. students in grades 8–12. Students are mainly from Title I schools or low-income communities, with trained volunteer tutors. We sampled two sessions from distinct tutor–student pairs. Audio and multimodal data were drawn from a school-embedded, one-on-one virtual tutoring program serving middle school students across the U.S. Approximately 80\% of students from low-income backgrounds and tutors consisting of college students. From this source, we analyzed two audio and two multimodal transcripts. Multimodal transcripts combined spoken dialogue with time-aligned logs of student interactions (e.g., mouse movements and interface actions), represented textually alongside utterances. All data were de-identified and segmented into ~200 utterances per modality.

\textbf{Codebook Development.} While prior LLM-assisted codebook work largely relies on GPT-based models, we used Gemini 2.5 Pro for its LearnLM training, which emphasizes instructional reasoning and outperforms GPT-4o on teaching-related tasks \cite{modi2024learnlm}. Following Barany et al.’s workflow \cite{barany2024chatgpt}, we designed a persona-based, step-structured prompt that frames the model as a sociocultural discourse researcher and anchors coding in Accountable Talk theory \cite{zambrano2026data}. To ground generation in data, Gemini was provided with 10 tutoring transcripts (amounting to 5 hours), which informed preliminary code construction prior to human refinement.  Gemini 2.5 received the following prompt, followed by the data:
    \\
    \\
    \textit{Act as a sociocultural researcher specializing in classroom discourse analysis. I am providing you with a transcript of a tutoring session. Your task is to perform an initial open coding to develop a preliminary codebook.}
    \\
    \textit{\textbf{Theoretical Lens}:
    Your coding should be informed by the Accountable Talk Framework, dialogic theories, and related perspectives. This means you need to look for specific, observable “talk moves” that demonstrate how both the tutor and the student contribute to accountable discourse.}
    \\
    \textit{\textbf{Codebook Requirements}:
    Identify and code both tutor and student talk moves. For each emergent 
    code, provide:• Code Name; • Category (drawn from theoretical lens); • Speaker Type; • Definition; • Two example quotes.}
    \\
    \\
To ensure prompt stability, we set temperature to 0 and ran the prompt three times on the same data in separate sessions \cite{barany2024chatgpt,hou2024prompt,zambrano2026data}. Following Liu et al.’s criterion, prompts were retained only if no more than two constructs differed across runs \cite{liu2025exploring}. Each run produced 8–12 candidate themes, which were then refined using Weston et al.’s iterative coding and refinement until convergence \cite{weston2001analyzing}.

\textbf{Coding Procedures.} Two expert annotators, blinded to the codebooks’ origin and unfamiliar with the datasets, completed a structured training and coding workflow. They first practiced by independently applying each codebook to chat, audio, and multimodal data, \cite{weston2001analyzing}. Then annotators met for a one-hour social moderation session to align interpretations  \cite{herrenkohl2013investigating}. Annotators then independently coded three new datasets spanning all modalities using a counterbalanced design to mitigate order effects. Coding was conducted via a structured interface with pre-populated dropdown menus, allowing multiple or no codes per utterance or “No Code” when applicable.

\begin{figure}[h!]
    \centering
    \includegraphics[width=0.90\textwidth]{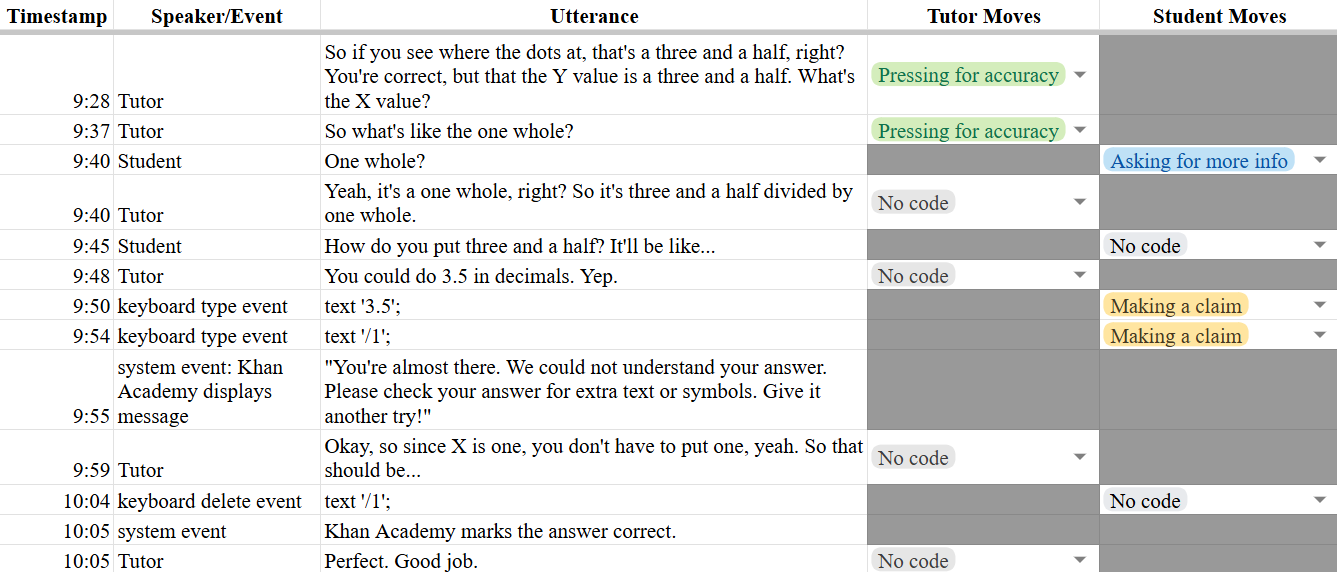} 
    \caption{Sample of multimodal dataset annotation interface, organized by speaker, utterance, and the corresponding drop-down to select tutor or student moves.}
    \label{fig:my_label}
\end{figure}
\vspace{-4 mm}
\textbf{Codebook Evaluation.} We evaluated both codebooks using a multi-method approach combining inter-rater reliability (IRR), perceived utility, and conceptual overlap analysis. Annotators were experts with over 20 years of teaching experience, held PhDs in the learning sciences, and were trained in qualitative coding. For RQ1, we assessed IRR across modalities and collected perceived utility via a survey adapted from Barany et al. \cite{barany2024chatgpt}, measuring ease of use, clarity, alignment, mutual exclusivity, exhaustiveness, and theoretical fidelity. For RQ2, we conceptually compared both codebooks, categorizing constructs as direct, functional, partial, or human-only overlaps based on their functions.
\vspace{-2 mm}
\section{Results}
\subsection{RQ1: Reliability and Utility of Codebooks and Data Modalities}

\textbf{\textit{Codebook Reliability Comparisons.}} Table~1 reports post-moderation Cohen’s $\kappa$ across chat, audio, and multimodal tutoring data. It shows moderate to substantial IRR for both codebooks, with \textit{TalkMoves} achieving higher average agreement ($\kappa$ = 0.74 vs. 0.64). Notably, four Accountable Talk codes (\texttt{Keeping everyone together}, \texttt{Getting students to relate to another’s ideas}, \texttt{Relating to another student}, and \texttt{Getting students to relate}) did not appear in the tutoring datasets, reflecting classroom-oriented constructs absent in one-on-one interactions, while all AI-human codes appeared at least once. Consistent with this, Figure~2 shows that the AI-human codebook produced more annotations across all modalities; for example, in audio, \textit{TalkMoves} captured 26 utterances compared to 41.5 for the AI-human codebook, suggesting that although the AI-human codebook achieved lower IRR, it captured a broader range of tutoring moves.

\begin{table}[t]
\centering
\footnotesize
\setlength{\tabcolsep}{3pt}
\renewcommand{\arraystretch}{0.95}
\caption{IRR between two experienced annotators for \textit{TalkMoves} and AI-human codebooks across tutoring data modalities. $\kappa > 0.6$ in bold (* denotes the code did not appear in the dataset for this modality).}
\label{tab:irr-kappa}

\begin{tabular}{l|l|c|c|c}
\hline
Codebook & Code & Chat ($\kappa$) & Audio ($\kappa$) & Multi ($\kappa$) \\
\hline

 & \texttt{Keeping together} & * & * & * \\
 & \texttt{Getting students to relate} & * & * & * \\
 & \texttt{Restating} & 0 & \textbf{0.8} & 0 \\
 & \texttt{Pressing for accuracy} & \textbf{0.88} & \textbf{1} & \textbf{0.82} \\
 & \texttt{Revoicing} & -0.01 & * & * \\
\textit{TalkMoves} & \texttt{Pressing for reasoning} & -0.01 & \textbf{1} & \textbf{1} \\
 & \texttt{Relating to student} & * & * & * \\
 & \texttt{Asking for more info} & \textbf{0.87} & * & 0.48 \\
 & \texttt{Making a claim} & \textbf{0.92} & \textbf{1} & \textbf{0.75} \\
 & \texttt{Providing evidence} & \textbf{0.8} & \textbf{1} & 0.49 \\
\cline{2-5}
 & Average across modality & \textbf{0.69} & \textbf{0.96} & 0.59 \\
\cline{2-5}
 & Average $\kappa$ & \multicolumn{3}{c}{\textbf{0.74}} \\
\hline

 & \texttt{Opening floor} & 0.39 & 0.58 & -0.01 \\
 & \texttt{Affirmation} & \textbf{0.71} & \textbf{0.92} & \textbf{0.92} \\
 & \texttt{Guided question} & \textbf{0.87} & \textbf{1} & \textbf{0.79} \\
 & \texttt{Error accountability} & 0.46 & \textbf{1} & \textbf{0.88} \\
AI-human & \texttt{Prompt thinkaloud} & -0.01 & \textbf{0.79} & -0.01 \\
 & \texttt{Help-seeking} & \textbf{0.94} & 0 & \textbf{0.65} \\
 & \texttt{Stating answer} & \textbf{0.84} & \textbf{0.78} & \textbf{0.67} \\
 & \texttt{Thinkaloud} & 0.48 & \textbf{0.83} & \textbf{0.85} \\
 \cline{2-5}
 & Average across modality & 0.59 & \textbf{0.74} & 0.59 \\
\cline{2-5}
 & Average $\kappa$ & \multicolumn{3}{c}{\textbf{0.64}} \\
\hline

\end{tabular}
\end{table}

\begin{figure}[h!]
    \centering
    \includegraphics[width=0.55\textwidth]{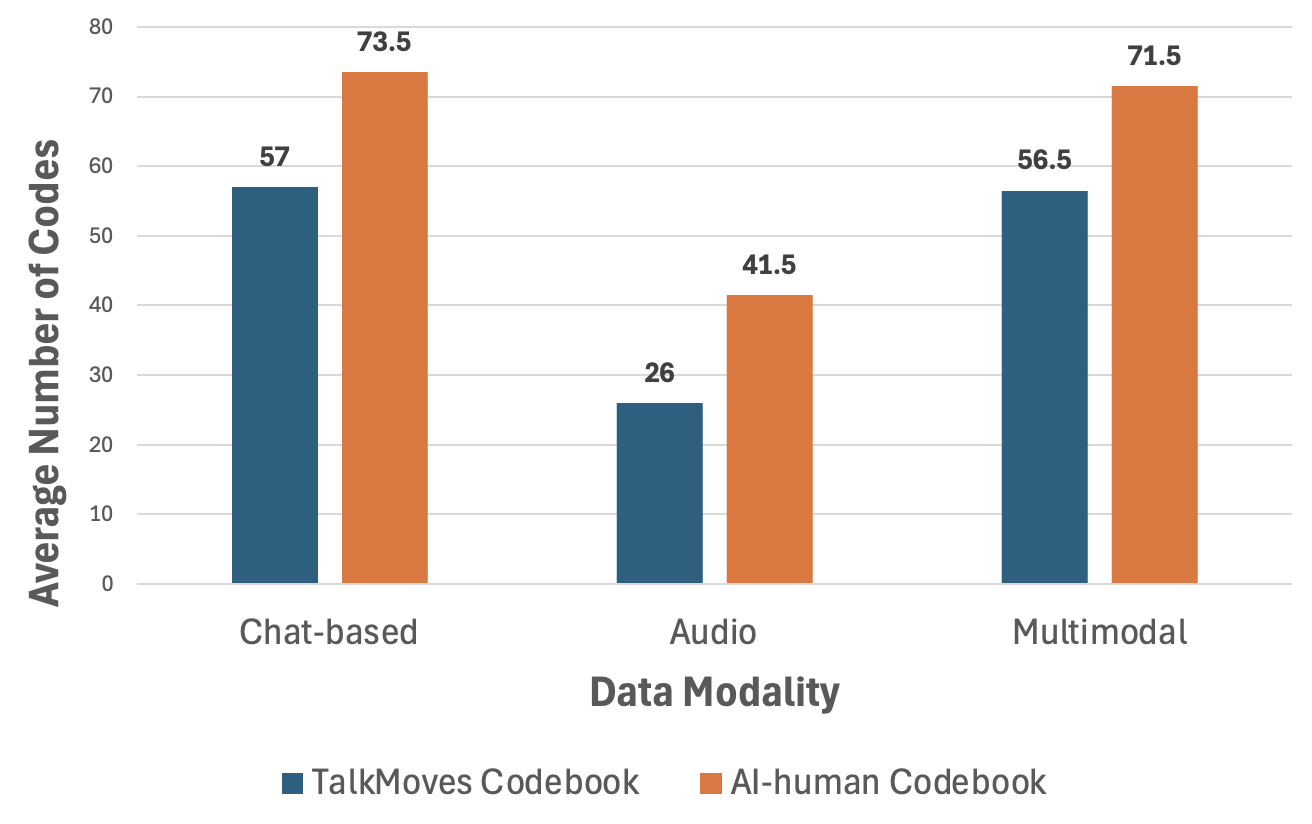} 
    \caption{Average number of codes identified using the human-developed \textit{TalkMoves} codebook (blue) and the AI-human codebook (orange) grouped by modality.}
    \label{fig:my_label}
\end{figure}

IRR and coverage varied by modality. Reliability was highest in audio (\textit{TalkMoves}: 0.96; AI-human: 0.74), equal in multimodal data (0.59), and lowest in chat-based interactions, particularly for constructs requiring extended reasoning (e.g., \texttt{Prompt thinkaloud}). At the same time, audio yielded fewer annotations overall, while chat and multimodal data increased coverage but introduced ambiguity. This pattern suggests a tradeoff: speech-based data supports more consistent coding, whereas multimodal and chat contexts expand observable interactions at the cost of lower agreement.

\textbf{\textit{Codebook Utility Comparisons.}} Table~2 summarizes annotators’ ratings across six utility dimensions. Overall, the hybrid AI-human codebook performed as well as or better than \textit{TalkMoves} across most categories. For example, both codebooks received equal ratings for \textit{Code clarity} (4.5 vs. 4.5), while the AI-human codebook scored higher on \textit{Ease of use} (4.75 vs. 4.5). However, it scored lower on \textit{Theory fidelity} (3.0 vs. 4.5), indicating a tradeoff between theoretical alignment and a more data-driven, context-sensitive approach. Exhaustiveness was rated low for both codebooks (1.25 vs. 1.75), with annotator feedback highlighting key gaps. For \textit{TalkMoves}, coders reported difficulty applying many classroom-oriented codes, often resorting to “No Code.” While the AI-human codebook better matched tutoring discourse, annotators still noted missing categories, such as checking student for understanding and direct explanation.
\vspace{-8 mm}
\begin{table}
\centering
\caption{Coder agreement ratings of codebook utility (1 = lowest; 5 = highest)}
\resizebox{\textwidth}{!}{%
\begin{tabular}{l|llllll} 
\hline
Codebook & Ease of use & Code clarity & Alignment & Mutual exclusivity & Exhaustiveness & Theory fidelity  \\ 
\hline
Human    & 4.5                                                         & 4.5                                                            & 4         & 5                                                                  & 1.25                                                            & 4.5                                                        \\ 
\hline
AI-human & 4.75                                                        & 4.5                                                            & 4.5       & 5                                                                  & 1.75                                                            & 3                                                          \\
\hline
\end{tabular}}
\end{table}
\vspace{-8 mm}
\subsection{RQ2: Theme Comparison Across Both Codebooks}

Table~3 compares overlapping and unique constructs, showing strongest alignment for core Accountable Talk functions and differences in context-specific orchestration and scaffolding moves. Of the 13 codes examined, three showed direct overlap across tutor and student roles, including practices related to making reasoning explicit (\texttt{Pressing for reasoning} vs. \texttt{Prompt thinkaloud}) and student explanation and help-seeking (\texttt{Providing evidence or reasoning} vs. \texttt{Thinkaloud}; \texttt{Asking for more info} vs. \texttt{Help-seeking}). Three constructs showed functional overlap, reflecting shared discourse purposes with differences in granularity. For example, \texttt{Pressing for accuracy} and \texttt{Guided procedural question} both prompt students to clarify or verify their thinking, though the former emphasizes conceptual precision while the latter reflects a stepwise approach typical of tutoring. Two constructs showed partial overlap, with shared intent but reduced scope in the AI-human codes. \texttt{Making a claim} emphasizes justification within collective reasoning, whereas \texttt{Stating an answer} captures a narrower response. Differences were concentrated in \textit{TalkMoves}-specific constructs (e.g., \texttt{Revoicing}, \texttt{Getting students to relate to another’s ideas}, \texttt{Relating to another student}), which reflect classroom-oriented peer interaction absent in one-on-one tutoring.

\begin{table}[t]
\centering
\footnotesize
\setlength{\tabcolsep}{4pt}
\renewcommand{\arraystretch}{0.95}
\caption{Categories across codebooks grouped by tutor and student moves}
\label{tab:my-table}

\resizebox{\textwidth}{!}{
\begin{tabular}{p{0.28\textwidth}|p{0.28\textwidth}|p{0.28\textwidth}|p{0.12\textwidth}}
\hline
\textbf{Function} &
\textbf{\textit{TalkMoves} Codes} &
\textbf{AI-human Codes} &
\textbf{Overlap*} \\ 
\hline

\multicolumn{4}{c}{\textbf{Tutor Moves}} \\ \hline
Making reasoning explicit &
\texttt{Pressing for reasoning} &
\texttt{Prompt thinkaloud} &
Direct \\ \hline

Eliciting student thinking &
\texttt{Pressing for accuracy} &
\texttt{Guided question} &
Functional \\ \hline

Directing attention to errors &
\texttt{Pressing for accuracy} &
\texttt{Error accountability} &
Partial \\ \hline

Keeping shared understanding &
\texttt{Keeping everyone together} &
\texttt{Opening floor} &
Functional \\ \hline

Responding to contributions &
\texttt{Restating} &
\texttt{Affirmation} &
Functional \\ \hline

Reformulating contributions &
\texttt{Revoicing} &
\texttt{—} &
Human only \\ \hline

Classroom orchestration &
\texttt{Getting students to relate} &
\texttt{—} &
Human only \\ \hline

\multicolumn{4}{c}{\textbf{Student Moves}} \\ \hline
Making reasoning explicit &
\texttt{Providing evidence} &
\texttt{Thinkaloud} &
Direct \\ \hline

Seeking support &
\texttt{Asking for more info} &
\texttt{Help-seeking} &
Direct \\ \hline

Self-expression &
\texttt{Making a claim} &
\texttt{Stating answer} &
Partial \\ \hline

Engaging with peers &
\texttt{Relating to another student} &
\texttt{—} &
Human only \\ \hline

\multicolumn{4}{p{\textwidth}}{%
\footnotesize
\textit{Note.} *Direct overlap indicates closely aligned discourse functions with different labels; Functional overlap indicates similar communicative purposes with differences in granularity; Partial overlap indicates overlapping intent with reduced scope; Human-only indicates constructs unique to the human codebook.
} \\

\end{tabular}
}
\end{table}

\section{Discussion \& Conclusion}
\vspace{-3 mm}
In this paper, we explored how Accountable Talk codebooks generalize to tutoring multimodal interaction data across reliability, utility, and interpretability.

\textbf{Reliability and code count varied systematically by data modality.} With respect to RQ1, coder agreement was highest for audio data and lower for chat-based and multimodal interactions. We hypothesize that this pattern reflects the fact that both codebooks were developed using audio transcripts, making them better aligned with speech-based interactional cues. Lower reliability in chat and multimodal data may also reflect annotators’ greater experience with audio transcript annotation. In contrast, audio datasets yielded substantially fewer annotations on average than chat-based and multimodal datasets. This suggests that while visual artifacts introduce interpretive ambiguity, they also create additional opportunities for annotation that are not expressed through speech alone. For example, in Figure~1, an annotator coded a student’s typed response following a tutor’s verbal prompt as \texttt{Making a Claim}, an interaction that would not be observable in audio-only data.

\textbf{The AI-human codebook demonstrated higher utility than the \textit{TalkMoves} codebook among expert annotators.} Attending to RQ1,  across utility dimensions, expert annotators consistently rated the hybrid AI-human codebook as more usable than the human-developed \textit{TalkMoves} codebook, except for explicit theory alignment. These findings align with prior work showing that hybrid human-AI workflows can produce codebooks that are more accessible to annotators, particularly when derived from  interactional data \cite{barany2024chatgpt}.

\textbf{The AI-human codebook preserved core Accountable Talk functions related to reasoning while adapting or omitting classroom discourse facilitation moves.} With respect to RQ2, this suggests that practices such as making thinking explicit and articulating explanations generalize well from classroom to tutoring contexts. In contrast, differences cluster around interactional scope and instructional context: classroom coordination moves in \textit{TalkMoves} are absent in the AI-human codebook, while tutoring-specific constructs emphasize procedural scaffolding and individualized guidance. Together, these findings highlight both the robustness and contextual limits of \textit{TalkMoves} beyond whole-class discourse.

\textbf{Limitations.} The empirical evaluation was conducted on a dataset of approximately 600 utterances, drawn from six tutoring sessions, limiting generalizability. In addition, the hybrid AI-human codebook was generated using a single LLM (Gemini 2.5 Pro), constraining conclusions about the transferability of this approach to other model architectures. Finally, both codebooks were developed using audio-only transcripts, which may have shaped the constructs identified and limited their sensitivity to modality-specific interactional features.

\textbf{Conclusion.} This study examined whether the \textit{TalkMoves} codebook generalizes to one-on-one tutoring across chat, audio, and multimodal data. Regarding RQ1, both codebooks achieved moderate to high reliability, with \textit{TalkMoves} showing higher agreement, especially in audio, while the AI-human codebook demonstrated higher utility and broader coverage. Reliability and code distribution varied by modality: audio supported more consistent coding, whereas chat and multimodal data increased coverage but introduced ambiguity. This highlights a tradeoff between theoretical consistency and contextual sensitivity. Together, these findings suggest it may be time to move beyond directly reusing \textit{TalkMoves} for tutoring annotation and instead develop modality-aware, tutoring-grounded codebooks. AI-human codebooks could be a promising path, but they introduce trade-offs in theory fidelity and reproducibility.

\section*{Acknowledgments}
This work was supported in part by the Learning Engineering Virtual Institute, the Gates Foundation, and the Chan Zuckerberg Initiative. The opinions, findings, and conclusions expressed are those of the authors.

\bibliographystyle{splncs04}
\bibliography{main} 

\end{document}